\documentclass[prd,nofootinbib,preprint,superscriptaddress]{revtex4-1}

\usepackage{amsmath, amssymb, amsthm, graphicx, epsfig, fancyhdr,epsfig}

\usepackage{tikz-feynman}
\tikzfeynmanset{compat=1.1.0}

\usepackage{tikzsymbols}

\usepackage{float}
\usepackage{xcolor}

\newcommand{\be}{\begin{equation}}
\newcommand{\ee}{\end{equation}}
\newcommand{\bea}{\begin{eqnarray}}
\newcommand{\eea}{\end{eqnarray}}

\begin{document}
\title{
Kination-like Era Driven by the Effective Inflaton/Higgs Potential
}

\author{Anish Ghoshal}
\email{anish.ghoshal@fuw.edu.pl}
\affiliation{Institute of Theoretical Physics, Faculty of Physics, University of Warsaw, \\ ul. Pasteura 5, 02-093 Warsaw, Poland}

\author{Nobuchika Okada}
\email{okadan@ua.edu}
\affiliation{Department  of  Physics  and Astronomy, \\ University  of  Alabama,  Tuscaloosa,  Alabama  35487,  USA}

\author{Arnab Paul}
\email{arnabpaul9292@gmail.com  }
\affiliation{Centre for Strings, Gravitation and Cosmology, Department of Physics, Indian Institute of
Technology Madras, Chennai 600036, India}

\begin{abstract}
\textit{
Based on the minimal $U(1)_X$ extended Standard Model,  we explore cosmic inflation where the $U(1)_X$ Higgs field serves as the inflaton. 
We demonstrate that a stiff era with an equation of state $w > 1/3$ can emerge during the inflaton's oscillatory phase after inflation, 
  driven by the Coleman-Weinberg potential of the inflaton, arising due to radiative corrections. 
This leads to significant modulation and enhancement of the irreducible stochastic gravitational wave (GW) background from inflation, 
  deviating from the conventional scale-invariant spectrum. 
Such a distinct GW spectrum could be detectable by next-generation GW interferometer missions, such as U-DECIGO. 
In our framework, the GW spectrum depends on the $U(1)_X$ gauge coupling and the mass of the $U(1)_X$ gauge boson ($Z^\prime$). 
As a result, future GW observations and $Z^\prime$ boson resonance searches at high-energy collider experiments 
   are complementary to one another. 
}
\end{abstract}

\maketitle

\newpage

\section{Introduction}


Cosmic inflation has been proposed to solve what is known as the horizon and flatness problems  
   in the Big Bang Cosmology \cite{Brout:1977ix,Sato:1980yn,Guth:1980zm,Linde:1981mu,Starobinsky:1982ee}. 
This scenario can also provide the origin of primordial density fluctuations for the seed of large-scale structure (LSS) formation and a primordial Stochastic Gravitational Wave Background (SGWB). 
Although the inflationary predictions are constrained by the precise measurements 
  of the Cosmic Microwave Background (CMB) anisotropy \cite{Planck:2018vyg}, 
  such as the scalar power spectrum, spectral index and tensor-to-scalar ratio \cite{Ade:2018gkx, Akrami:2018odb,BICEP:2021xfz}, 
  the evolution of the universe between the end of inflation and the onset of Big Bang Nucleosynthesis (BBN) 
  remains unknown \cite{Allahverdi:2020bys}. 
In the standard Big Bang Cosmology, the post-inflationary universe is assumed to be radiation-dominated, 
  and the energy density of the universe is dominated by radiation whose equation of state (EOS) is 
  $w = 1/3$, 
  where $w = p/\rho$ with $\rho$ and $p$ is the energy density and the pressure of the state, respectively. 
  
Alternative to the standard cosmological history with $w=1/3$, several cosmological scenarios, for example, 
   the dynamics of the inflaton at the end of inflation can trigger a so-called {\it stiff } EOS, which means $w > 1/3$,
   such that the total energy density of the universe redshifts faster than radiation    
   \cite{Spokoiny:1993kt,Joyce:1996cp,Ferreira:1997hj,Bernal:2020ywq,Chen:2024roo,Saha:2020bis,Ghoshal:2022ruy}. 
Such scenarios of the early universe evolving with $w> 1/3$ (referred henceforth as "kination-like" era) in the post-inflationary universe may impact an SGWB of primordial origin, such as inflation, reheating/preheating, first-order phase transitions, and cosmic strings (see Ref.~\cite{Caprini:2018mtu} for a review).  
It is particularly interesting that the kination-like or stiff dominated era leads to a significant growth of tensor perturbation modes 
  at large frequencies and gravitational wave detectors could observe enhanced SGWBs 
  in the very near future 
   \cite{ Giovannini:1998bp, Giovannini:2009kg, Riazuelo:2000fc, Sahni:2001qp, Seto:2003kc, Tashiro:2003qp, Nakayama:2008ip, Nakayama:2008wy, Durrer:2011bi, Kuroyanagi:2011fy, Kuroyanagi:2018csn, Haque:2021dha,Jinno:2012xb, Lasky:2015lej, Li:2016mmc, Saikawa:2018rcs, Caldwell:2018giq, Bernal:2019lpc, Figueroa:2019paj, DEramo:2019tit,Li:2013nal, Bernal:2020ywq, Odintsov:2021kup, Odintsov:2021urx, Li:2016mmc,Li:2021htg,Dimopoulos:2022mce,Co:2021lkc,Ghoshal:2022ruy,Vagnozzi:2020gtf,Benetti:2021uea,Hook:2020phx} 
   (see Ref.~\cite{Gouttenoire:2021jhk} for a review).

In this paper, we investigate a possible underlying physics of Beyond the Standard Model (BSM) 
 that can realize the kination-like era in the early universe, 
   in which the end of inflation is not followed by the standard Bing Bang era but instead by the kination-like era with the stiff EOS of $w >1/3$.  
Based on the minimal $U(1)_X$ extension of the SM, we will show how the kination-like era can be realized after the end of inflation, 
   which is driven by the $U(1)_X$ Higgs field, 
   leading to large detectable SGWB signals due to the amplification of energy density of primordial GWs. 
The resultant SGWB spectrum correlates to the $U(1)_X$ gauge coupling ($g_X$) value 
  and the  $U(1)_X$ gauge boson ($Z^\prime$) mass ($M_{Z^\prime}$),
  and this correlation provides us with the complementarity of our cosmological scenario with the kination-like era 
  to the low energy physics, particularly the $Z^\prime$ resonance search at the current and future Large Hadron Collider (LHC) experiments.

Motivations for considering a U(1)$_X$ extension of the SM stem from the need to explain the tiny neutrino masses and neutrino flavor mixing 
   observed through the neutrino oscillation phenomena \cite{ParticleDataGroup:2020ssz}. 
The seesaw mechanism \cite{Minkowski:1977sc,Yanagida:1979as,Gell-Mann:1979vob,Mohapatra:1979ia,Schechter:1980gr} 
   is a well-known mechanism in order to naturally generate the tiny neutrino Majorana masses, 
   where the SM is extended by introducing SM-gauge singlet right-handed neutrinos (RHNs) with their Majorana mass terms.  
 The minimal $U(1)_{B-L}$ extended SM  \cite{Davidson:1979wr,Mohapatra:1980qe,Mohapatra:1980qe,Marshak:1979fm,Davidson:1978pm,Davidson:1987mh,Wetterich:1981bx,Masiero:1982fi,Mohapatra:1982xz,Buchmuller:1991ce} 
   is a simple extension of the SM where the global $B-L$ (baryon minus lepton number) symmetry, 
   which the SM accidentally possesses, is gauged. 
This gauged $B-L$ symmetry extension requires the introduction of three RHNs to cancel all $U(1)_{B-L}$ related anomalies. 
The $U(1)_{B-L}$ gauge symmetry breaking by a $B-L $ Higgs generates not only $Z^\prime$ boson mass, 
  but also Majorana masses for the RHNs, and hence the seesaw mechanism is automatically incorporated. 
The minimal U(1)$_X$ extended SM is a generalization of the minimal $U(1)_{B-L}$ model 
  by assigning the $U(1)_X$ charge of a particle as a linear combination of the SM hypercharge ($U(1)_Y$) 
   and the $U(1)_{B-L}$ charge \cite{Appelquist:2002mw, Oda:2015gna}. 
Except for the generalization of $U(1)_X$ charge assignment, the structure of the minimal U(1)$_X$ model is similar to that of the minimal $B-L$ model. 
The minimal $U(1)_X$ model offers various interesting phenomenology, such as\footnote{
There is extensive literature on these topics; however, we focus our citations on a few seminal papers in each area.
} 
  a resolution of the SM vacuum instability \cite{Das:2015nwk, Oda:2015gna, Das:2016zue}, 
  dark matter (DM) physics with one RHN identified as a DM \cite{Anisimov:2008gg, Okada:2010wd}, 
  complementarities between LHC physics and RHN DM physics \cite{Okada:2016gsh, Okada:2016tci}, 
  RHN productions at future high energy colliders \cite{Das:2017flq, Das:2017deo, Das:2018tbd}, 
  a cosmic inflation scenario with identification of the $U(1)_X$ Higgs field as inflaton \cite{Okada:2011en, Okada:2016ssd} and formation of Primordial Black Hole and scalar-induced Gravitational Waves \cite{Ghoshal:2024hfk}.

This paper is organized as follows: 
In section \ref{sec:2}, we describe the $U(1)_X$ extended SM model.  
In section \ref{sec:3}, we discuss the inflation scenario by identifying the $U(1)_X$ Higgs field as inflation 
  and how the kination-like era is realized after inflation. 
In section IV, we discuss the complementary between SGWB detections and LHC physics for $Z^\prime$ boson. 
Section IV is devoted for conclusion and discussion.

\section {The minimal $U(1)_{X}$ Model}
\label{sec:2}
\begin{table}[t]
\begin{tabular}{|c|cccc|}
\hline
  &~~~$SU(3)_c$&~~~$SU(2)_L$&~~~$U(1)_Y$~~~&~~~$U(1)_X$\\
  \hline
  $q_L^i$  & ${\bf 3}$ & ${\bf 2}$ & $\frac{1}{6}$ & $\frac{1}{6} x_H+ \frac{1}{3} x_\Phi$ \\ 
  $u_R^i$ & ${\bf 3}$ & ${\bf 1}$ & $\frac{2}{3}$ & $\frac{2}{3} x_H+ \frac{1}{3} x_\Phi$ \\
  $d_R^i$ & ${\bf 3}$ & ${\bf 1}$ & $ - \frac{1}{3} $ & $ -\frac{1}{3} x_H+ \frac{1}{3} x_\Phi$ \\
  $\ell_L^i$ & ${\bf 1}$ & ${\bf 2}$ & $ -\frac{1}{2} $ & $ -\frac{1}{2}  x_H - x_\Phi$ \\
  $e_R^i$ & ${\bf 1}$ & ${\bf 1}$ & $-1$ &$ -x_H -x_\Phi$ \\
  $H$ & $ {\bf 1}$ & ${\bf 2}$ & $ \frac{1}{2}$  & $\frac{1}{2}  x_H$ \\
\hline
  $N_R^i$&${\bf 1}$&${\bf 1}$ & $0$ &  $ -x_\Phi$ \\
  $\Phi$ & ${\bf 1}$&${\bf 1} $ & $0$ & $2x_\Phi$ \\
  \hline
\end{tabular}
\caption{\it
Particle contents of the minimal $U(1)_X$ model. 
The index $i=1,2,3$ denotes the generations, and $x_H$ and $x_\Phi$ are real parameters 
 that are not fixed by anomaly cancellation conditions.
By redefining the $U(1)_X$ gauge coupling, we set $x_\Phi=1$ without loss of generality (for $x_\Phi \neq 0$). 
}
\label{tab:Contents}
\end{table}

The model is based on the gauge group $SU(3)_c\times SU(2)_L\times U(1)_Y\times U(1)_X$, and  
  the SM gauge group is extended with an extra $U(1)_X$.
The particle contents of the minimal $U(1)_X$ model are summarized in TABLE \ref{tab:Contents}. 
In addition to the SM particles, three RHNs ($N_R^i, i=1,2,3$) and one complex scalar $\Phi$ are introduced; 
  the former is necessary for the theory to be free from gauge and mixed-gauge and gravitational anomalies, 
  and the latter is responsible for breaking the $U(1)_X$ symmetry and generating Majorana masses of the RHNs. 
As shown in TABLE \ref{tab:Contents}, $U(1)_X$ charges for particles are determined by only two real parameters $x_H$ and $x_\Phi$, 
  which are not determined by the anomaly cancellation conditions. 
As $x_\Phi$, if it is non-zero, may be absorbed into redefinition of the $U(1)_X$ gauge coupling ($g_X$), 
  we shall set $x_\Phi=1$ throughout the paper, leaving the remaining $x_H$ as a unique free parameter for the charge assignment. 
The minimal $U(1)_{B-L}$ model is realized as a special case with $x_H=0$, 
  whereas $x_H= -4/5$ and $x_H=4$ correspond to the standard and flipped $SU(5)\times U(1)_X$ grand unified theories (GUTs) 
  which may be embedded in the $SO(10)$ GUT.

Since the abelian $U(1)_Y$ and $U(1)_X$ gauge bosons generally have a kinetic mixing, 
   the $U(1)_Y \times U(1)_X$ gauge couplings may be organized into the form of a triangular matrix
   with a mixed gauge coupling $\widetilde g_1$ \cite{delAguila:1988jz} , 
   and the covariant derivatives of the model are given by
\begin{equation}\label{eqn:CovDeriv}
  D_\mu= \partial_\mu-ig_3T^\alpha G^\alpha_\mu-ig_2 T^a W_\mu^a 
  -ig_1 Y B_\mu-i(\widetilde g_1 Y+g_X X)Z_\mu^\prime,
\end{equation}
where $G^\alpha_\mu$, $W^a_\mu$, $B_\mu$, $Z_\mu^\prime$ are the gauge fields of 
  $SU(3)_c$, $SU(2)_L$, $U(1)_Y$ and $U(1)_X$, respectively, 
  $g_3$, $g_2$, $g_1$, $g_X$ are the corresponding gauge couplings, 
  and $T^\alpha$, $T^a$, $Y$, $X$ are the corresponding generators and charges. 
Even if we set $\widetilde g_1=0$ at an energy scale, quantum corrections generate $\widetilde g_1 \neq 0$ at another scale. 
However, it can be shown that the effect of the flow of the Renormalization Group (RG) 
  for $|\widetilde g_1/g_X|$ is at most a few percent (see, for example, Ref.~\cite{Okada:2017dqs}), 
  and we ignore $\widetilde g_1$ in the following analysis.

The Yukawa terms of the SM are extended to include 
\begin{equation}
  {\cal L}_{\rm Yukawa} \supset
  -y_D^{ij}\overline{\ell_L^i} \widetilde H N_R^j
  -\frac{1}{2} y_M^i\Phi\overline{N_R^{ic}}N_R^i+\text{h.c.}, 
\end{equation}
where $y_D$ and $y_M$ are the Dirac and Majorana Yukawa matrix, respectively, $\widetilde H\equiv i\sigma_2 H^*$, and we work on the basis of diagonal Majorana Yukawa matrix $y_M$.
When $\Phi$ develops a vacuum expectation value (VEV), $\langle \Phi \rangle =\phi_{min}/\sqrt{2}$,  
  the $U(1)_X$ gauge symmetry is broken and the $Z^\prime$ boson mass is generated: 
  $M_{Z^\prime} = 2 g_X \phi_{min}$
From the last two term of the above Yukawa interactions, Majorana mass terms for $N_R^i$ are also generated, 
  so that the seesaw mechanism generates small SM neutrino masses once the electroweak symmetry is broken. 
This fact justifies the minimal $U(1)_X$ model as a well-motivated extension of the SM. 
Baryogenesis via leptogenesis \cite{Fukugita:1986hr} is also readily incorporated. 
In the following discussion, we identify the $U(1)_X$ Higgs field $\Phi$ with infalton.

\section{Cosmic History before Radiation Domination}
\label{sec:3}
\subsection{Inflation via $U(1)_X$ Higgs with non-minimal coupling to gravity}
As mentioned above, we identify the $U(1)_X$ Higgs field $\Phi$ with inflation. 
We introduce a non-minimal coupling to gravity of the scalar field and write the Lagrangian in the Jordan frame 
   relevant to our inflation discussion as 
   (in Planck units where the reduced Planck mass of $M_P=2.4 \times 10^{18}$ GeV is set to be 1)
\begin{equation}\label{LJ}
{\cal L}_J = \frac{1}{2} \sqrt{-g_J} \left[ (1+\xi \phi^{2}) R_J - g_J^{\mu\nu} \partial_\mu \phi \partial_\nu \phi - 2 V_J(\phi) \right],
\end{equation}
where the index $J$ corresponds to quantities in the Jordan frame, $\xi$ is the non-minimal gravitational coupling, and $\phi = \sqrt{2} \Phi$ is a real scalar in the unitary gauge, which is identified with inflaton.    
Using  the conformal transformation, $g_E^{\mu\nu} = (1+\xi \phi^{2})^{{-1}}  g_{J}^{\mu\nu}$, 
  the Lagrangian in the Einstein frame is expressed as 
\begin{equation}\label{LE}
{\cal L}_E = \frac{1}{2} \sqrt{-g_E} \left[ R_E - g_E^{\mu\nu} \partial_\mu \varphi \partial_\nu \varphi - 2 V_E(\phi(\varphi)) \right], 
\end{equation}
where the inflaton in the Einstein frame ($\varphi$) is related to $\phi$ by 
\begin{equation}\label{trans}
\left( \frac{d\varphi}{d\phi} \right)^2 = \frac{1+\xi \phi^2 + 6\xi^{2} \phi^2}{(1+\xi \phi^2)^2}, 
\end{equation}
and the inflaton potential in the Einstein frame is given by
\begin{equation}\label{eframe}
V_E(\phi) = \frac{V_J(\phi)}{(1+\xi \phi^2)^2}. 
\end{equation}

We consider a renormalizable Higgs potential for $\phi$ (see the next subsection for details) 
  and the $U(1)_X$ symmetry is broken at an energy scale much lower than the inflation energy scale.  
Since the scale of $\phi$ during inflation is (trans-)Planckian, 
  the inflation potential is approximately given by $V_J(\phi)=\frac{1}{4}\lambda_\phi \phi^4$ 
  at the tree-level.    
In the inflationary analysis of the system (see, for example, \cite{Okada:2014lxa}), 
  once $\xi$ is fixed, the values of $\varphi_{start}$ and $\varphi_{end}$ are fixed 
  by considering the slow-roll violation at the end of inflation and enough e-foldings from the start to end. 
We use $\lambda_\phi$ at the tree-level in our inflationary analysis, 
  although the effective Higgs potential after taking 
  quantum corrections to $\lambda_\phi$ into account is crucial for realizing the kination-like era, 
  as we will discuss below.   
Since the difference between $\varphi_{start}$ and $\varphi_{end}$ is about an order of magnitude, 
  $\lambda_\phi$ is indeed almost constant during inflation. 
Note that Eq.~(\ref{eframe}) indicates that the quartic inflaton potential in the Jordan frame 
  becomes flat in the Einstein frame during inflation ($\xi \phi^2 \gtrsim 1$) 
  and the inflaton potential is nicely compatible with the slow-roll conditions. 
After the inflation, the inflaton amplitude reduces to $\xi \phi^2 < 1$, and the fields $\phi$ and $\varphi$ become identical with each other. 


\subsection{Post-Inflationary dynamics of inflaton}

It is known that for a potential $V(\phi) \propto \phi^n$, the EOS corresponding to the energy density of a oscillating scalar field 
   is given by $w \simeq \frac{n-2}{n+2}$\footnote{There may be corrections arising due to fluctuations of the $\phi$ and perturbations of the FRW metric, but unegligible for our case, see Ref.\cite{Cembranos:2015oya} for details.}, where $n$ is an integer.  
Hence, the EOS for $V(\phi) = \frac{1}{4}\lambda_\phi \phi^4$ theory with a constant $\lambda_\phi$ is $w=\frac{1}{3}$, 
  and the stiff EOS is obtained if a scalar potential is steeper than $\phi^4$, or equivalently, $n>4$. 
In our scenario, after the end of inflation, the inflaton amplitude reduces to $\xi \phi^2 < 1$ 
  and the inflaton starts oscillating along its potential of $V=\frac{1}{4}\lambda_\phi \phi^4$. 
Note that quantum corrections modify this tree-level potential to involve a term like $\sim \ln(\phi)\phi^4$, 
  so that the inflaton oscillates along the the effective inflaton potential with a slope steeper than $\phi^4$. 
Therefore, we expect that the kination-like era ($w > 1/3$) is realized during this oscillation  
  as long as the amplitude of the oscillation is sufficiently large until the inflaton is trapped 
  to oscillate around a potential minimum.

In our minimal $U(1)_X$  model, we consider the classically conformal system for the $U(1)_X$ Higgs sector,  
  which means that the model is defined as a massless theory at the potential origin 
  by imposing the massless renormalization condition for $\phi$ at $\phi=0$. 
Coleman and Weinberg first considered such a system and showed that $U(1)$ gauge symmetry is radiatively broken \cite{Coleman:1973jx}. 
Following Ref.~\cite{Coleman:1973jx},  we set $\lambda_\phi,  |Y_N^i|^2 \ll g_X^2$ and the effective $U(1)_X$ Higgs potential at the one-loop level 
  (Coleman-Weinberg potential) is found to be 
\begin{eqnarray}\label{Veff}
   V_{\rm eff} \simeq  \frac{1}{4} \left( \lambda_\phi +  \frac{\beta_\lambda}{2} \ln\left[ \frac{\phi^2}{\phi_{min}^2}\right]\right) \phi^4 
      - \frac{1}{4}\lambda_\phi \phi_{min}^4, 
\end{eqnarray}
where $\beta_\lambda=\frac{1}{16\pi^2}96g_{X}^4$, and  the renormalization scale is set to be $\phi_{min}$ being the VEV of $\phi$ 
   at the potential minimum.  
The quartic coupling $\lambda_\phi$ has a relation to $g_X$,  $\lambda_\phi = \beta_\lambda/4$, 
   which is determined by the stationary condition, $dV_{\rm eff}/d \phi|_{\phi=\phi_{min}}=0$. 
Since the $U(1)_X$ Higgs field has no direct coupling wth the SM fermions, the effective potential is independent of $x_H$. 
In the above discussion, we also have set a mixed quartic coupling between the SM and $U(1)_X$ fields, 
  $\lambda_{mix} (H^\dagger H)(\Phi^\dagger \Phi)$, to be small, $|\lambda_{mix}| \ll g_X^2$, 
  and its contribution to $V_{\rm eff}$ is negligible. 
In our analysis, along with $\xi$, we use $\phi_{min}$ as a free parameter. 
Once these parameters are fixed, $g_X$ is determined from the inflationary analysis.

\subsubsection{Numerical Study with a Toy Model}

Given a potential, we can understand the dynamics of the inflaton by solving its equation of motion (EOM): 
\begin{eqnarray}\label{eom}
  \ddot\varphi+3\mathcal{H}\dot{\varphi}+V_{,\varphi}&=&0,
\end{eqnarray}
where the dot denotes derivative with respect to cosmic time t, $\mathcal{H}=\dot{a}/a$ is the Hubble parameter with $a$ being the scale factor of the Friedmann-Lemaitre-Robertson-Walker metric, and $V_{,\varphi}=\frac{dV}{d\varphi}$.
Once the EOM is solved and $\varphi(t)$ is obtained, the EOS, which is in general a function of time,  
  can be calculated from the evolution of energy density of the system 
  $\rho(t)=\frac{1}{2}\dot{\varphi}^2+V(\varphi)\varpropto a^{-3(1+w(t))}$. 
Due to the inflaton oscillation, $w(t)$ exhibits an oscillation behavior.  
To get a smooth EOS as a function of time, we calculate $w(t)$ as an average over a time interval spanning several oscillations.

\subsubsection{Semi-Analytic Calculations}

For the effective potential, solving the EOM from the inflation era to the oscillation era to evaluate $w(t)$ is numerically difficult. 
Toward a simpler but sufficiently accurate calculations, we develop a semi-analytic calculation (SAC) in this section. 

\begin{figure}[H]
\begin{center}
\includegraphics[height=0.45\textwidth]{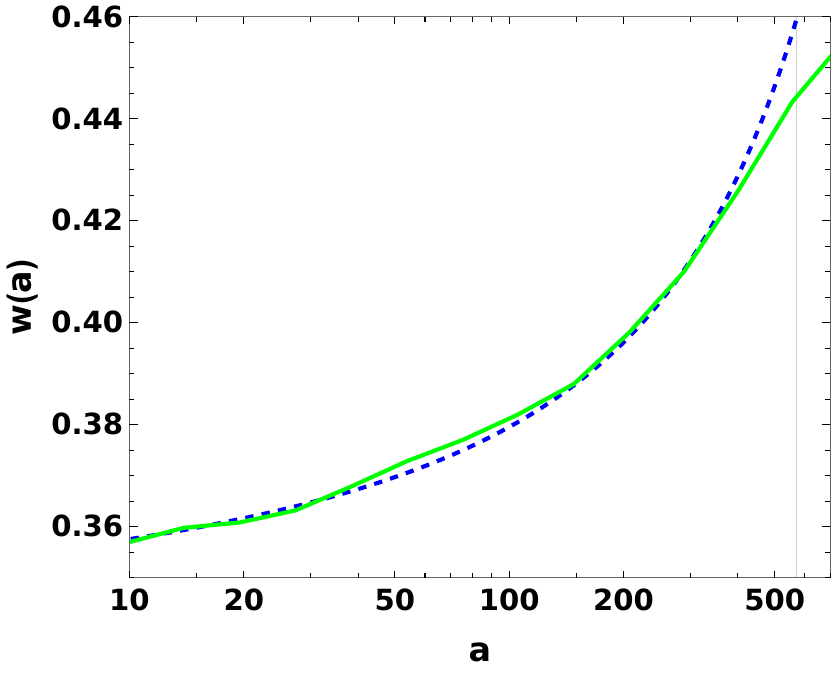}
 \caption{\it 
 $w$ versus $a$ from the numerical analysis (green) and from the SAC (blue dashed) for a toy model with potential given by $V_{\rm eff}$\ref{Veff}. Note that the SAC begins to deviate from the numerical result as the energy density of the field approaches the height of the potential at $\phi=0$ (given by the vertical gray line).
 }
\label{fig:potinf}
\end{center}
\end{figure}

As mentioned in the previous subsection, we calculate $w(t)$ as an average over a time interval spanning several oscillations 
   ($w_{avg}$) due to the oscillation behavior of $w(t)$. 
During the inflaton oscillation, if $w_{avg}$ evolves with time due to change of an effective exponent of the potential, 
   which we define as $p(\varphi)=\frac{d(\log(V(\varphi)))}{d\log(\varphi)}$, for a change of scale factor $\Delta a$, 
   the evolution of inflaton energy density is given by
\begin{equation}
    \frac{\rho(a+\Delta a)}{\rho(a)}\propto     \left(\frac{a+\Delta a}{a}\right)^{-3(w_{avg}+1)}
\end{equation}
Since $w=w_{avg}=\frac{p-2}{p+2}$ for a potential $V \propto \varphi^p$ with a constant $p$, 
   we may write $w_{avg}=\frac{p(\varphi)-2}{p(\varphi)+2}$, where 
   \begin{equation}
    p(\varphi)\sim\frac{\log (V(\varphi+\Delta\varphi))-\log(V(\varphi))}{\log (\varphi+\Delta\varphi)-\log(\varphi)}
    \label{pasphi}
\end{equation}   
for  a potential expressed as $V\sim\varphi^{p(\varphi)}$. 
Now, considering $\rho=V(\varphi)\sim\varphi^{p(\varphi)}$ and  $\rho\varpropto a^{-3(w_{avg}+1)}$, 
   a simple calculation leads to the following relation: 
\begin{equation}
    \Delta(\log(a))=(1+\Delta \log(\varphi))^{-\frac{p+2}{6}}-1\label{aasphi}.
\end{equation}
Eq.~(\ref{pasphi}) (connecting $p$ to $\varphi$) along with Eq.~(\ref{aasphi}) (connecting $\Delta a$ to $\Delta \varphi$) 
   enables us to calculate $w_{avg}$ as a function of $a$.
To check the accuracy of the SAC results, we consider a toy model and numerically evaluate $w(t)$. 
This result is to be compared with the result from the SAC. 
This comparison is shown in Fig.~\ref{fig:potinf}. 
The SAC result is consistent with the numerical result with a very good accuracy.

\subsection{Reheating after inflation}
The amplitude of oscillating inflaton dies down by the red-shift, and the inflaton is eventually trapped in one of the potential minima. 
Around the minimum, the inflaton potential is approximately quadratic and the evolution of the Universe behaves 
  as a matter-dominated era ($w=0$). 
This era continues until the inflaton decays to the SM particles and then, the radiation-dominated era of the standard Big-Bang cosmology begins. 
The reheating temperature $T_{rh}$ at the beginning of the Big-Bang Cosmology is determined by the decay width (or lifetime ) of the inflaton. 
The reheating temperature is essentially a free parameter of our scenario, and we are left with a specific $T_{rh}$ value 
   for a viable cosmological scenario.

The reheating temperature is related to e-folding number. 
A co-moving mode $k$ corresponding to the CMB ($k=0.05\rm Mpc^{-1}$) exits the Hubble horizon during inflation 
  with $k=a_k H_k$, where $a_k$ and $H_k$ corresponds to the scale factor and Hubble scale at horizon exit of the mode $k$. 
The co-moving Hubble horizon $(aH)^{-1}$ shrinks exponentially until end of inflation and then grows again letting the frozen modes 
  reenter the co-moving Hubble horizon. 
As the composition of the standard $\Lambda$CDM model is well constrained, 
  the evolution of Hubble in standard Big-Bang cosmology is given by $H^{late}=H_0(\Omega_{cdm}^0 a^{-3}+\Omega_b^0 a^{-3}+\Omega_r^0 a^{-4}+\Omega_\Lambda)$, where `$0$' signifies current epoch $a=a_0=1$, with the other notation used as in standard practice. 
Using $k=a_k^{late}H_k^{late}$, we can find the epoch of reentry of the co-moving mode $k$ at $a=a_k^{late}=4.5\times10^{-5}$.

Now, we express $k=a_k H_k$ as follows: 
\begin{equation}
    \ln\left(\frac{k}{a_k H_k}\right)=\ln\left(\frac{a_{end}}{a_k}\frac{a_{kin}^j}{a_{end}}...\frac{a_{kin}^i}{a_{kin}^{i-1}}\frac{a_{EMD}}{a_{kin}^i}\frac{a_{rh}}{a_{EMD}}\frac{a_k^{late}}{a_{rh}}\frac{k}{a_k^{late} H_k}\right)=0.
    \label{eq:k=aH}
\end{equation}
Here, the suffixes $end$, $kin$, $EMD$, and $rh$ correspond to the end of inflation, kination-like era, 
  early matter-domination and reheating, respectively.
This separation according to scale factors have been made in such a way that the EOS is approximately constant during each phase. 
The kination-like phase is expressed in steps like $\frac{a_{kin}^j}{a_{end}}...\frac{a_{kin}^i}{a_{kin}^{i-1}}\frac{a_{EMD}}{a_{kin}^i}$, 
  because the EOS evolves during that phase as we have discussed. 
Now, during some phase from $a_j$ to $a_i$ with EOS $w$, 
   we get $\ln\left(\frac{a_i}{a_j}\right)=-\frac{1}{3(w+1)}\ln\left(\frac{\rho(a_i)}{\rho(a_j)}\right)$. 
As we have already discussed, the evolution of universe is known from $a_k$ to $a_{EMD}$ and from $a_k^{late}$ to current epoch. 
We also know that the e.o.s of the phases from $a_{EMD}$ to $a_{rh}$ is 0 and from $a_{rh}$ to $a_k^{late}$ is $\frac{1}{3}$. 
This lets us find $a_{rh}$ and hence $T_{rh}$ from eq.\ref{eq:k=aH}, thereby fixing the duration of EMD phase. 

\medskip

\section{Gravitational Wave and Collider Signatures}
\subsection{Modulated GW spectrum}\label{mod}

The GW spectrum produced during the inflation is flat (scale-invariant). 
However, when the modes of the GW reenter the horizon during different epochs with different EOSs, 
  the flat GW spectrum gets modulated. 
The GW spectrum at present is given by

\begin{widetext}
\begin{equation}
	\Omega^{0}_{\rm GW} (k) = \Omega_{\rm GW}^{\rm 0, flat} \begin{cases}
		 1\,, & k < k_\text{rh} \\ 
		\left(\frac{k}{k_\text{rh}}\right)^{\frac{2(3w_{EMD}-1)}{1+3w_{EMD}}}\,, & k_\text{rh} \leq k \leq k_{\rm kin}^i \\
		\left(\frac{k_\text{kin}^i}{k_\text{rh}}\right)^{\frac{2(3w_{EMD}-1)}{1+3w_{EMD}}}\left(\frac{k}{k_\text{kin}^i}\right)^{\frac{2(3w_{kin}^i-1)}{1+3w_{kin}^i}}\,,& k_\text{kin}^i \leq k \leq k_\text{kin}^{i-1} \\
  ...\\
  \left(\frac{k_\text{kin}^i}{k_\text{rh}}\right)^{\frac{2(3w_{EMD}-1)}{1+3w_{EMD}}}\left(\frac{k_\text{kin}^{i-1}}{k_\text{kin}^i}\right)^{\frac{2(3w_{kin}^i-1)}{1+3w_{kin}^i}}...\left(\frac{k_{end}}{k_\text{kin}^j}\right)^{\frac{2(3w_{kin}^j-1)}{1+3w_{kin}^j}}\,,& k_\text{kin}^j \leq k \leq k_{end}
  \\
		0 &  k_{\rm end}<k
			\end{cases} \,
	\label{eq:GWspecTot}
\end{equation}
\end{widetext}
where the modes $k_i$ corresponds to the reentering mode at $a=a_i$ with $k_i=a_i H_i$.

\begin{equation}\label{eq:GW}
\Omega_{\rm GW}^{\rm 0, flat} = \frac{\Omega_{\gamma}^{0}}{24}  \left(\frac{g_{s,\rm eq}}{g_{s,k}}\right)^{\frac{4}{3}} \left(\frac{g_{k}}{g_{\gamma}^{0}} \right) \frac{2}{\pi^{2}} \frac{H_{\rm end}^{2}}{M_P^{2}}\,,
\end{equation}
where $g_{\gamma}^{0} = 2$ and $g_{k}$ is the d.o.f during the reentry of mode $k$. 
The frequency $f$ corresponding to a wave-number $k$ is related via $f=c\frac{k}{2\pi}$, where $c$ is the speed of light. $H_{\rm end}$ is chosen to be that maximal allowed $r=A_t/A_s<0.035$ by PLANCK where the tensor amplitude $A_t$ corresponds to $A_t=\frac{2}{\pi^2}\frac{H_{\rm end}^2}{M_P^2}$. We have assumed $\Omega^{0}_{\rm GW} (k)=0$ for modes which are sub-hubble at all times ($k>k_{end}$)\footnote{Actually $\Omega^{0}_{\rm GW} (k)\sim k^4$ for ($k>k_{end}$) when no regularisation has been taken into account. However if regularisation is done properly, the spectral energy density falls exponentially with $k$ \cite{Pi:2024kpw}.}.

\begin{table}[t]
 \begin{tabular}{| c | c | c |c|| c | c| c|} 
 \hline
 BM &$\xi$ & $N_{inf}$ & $\varphi_{min}$ (GeV) & $g_X$ & $M_{Z'}$ (GeV) & $T_{rh}$ (GeV) \\ [0.5ex] 
 \hline
 1  & 100 & 55 &  1.6$\times10^{9}$ & 0.0179 & 5.67$\times 10^{7}$ & 1.11 $\times 10^{7}$\\
 \hline
 2  & 100 & 55 &  5$\times10^{4}$ & 0.0161 & 1607 & 141.31\\
 \hline
 3  & 100 & 55.6 &  1.6$\times10^{9}$ & 0.0178 & 5.64$\times 10^{7}$ & 2.15 $\times 10^{7}$\\
 \hline
 4  & 100 & 55.3 &  5$\times10^{4}$ & 0.0160 & 1603 & 534.74\\
 \hline
 \end{tabular}
\caption{\it 
\it Benchmark points.}
\label{t1}
\end{table} 
As mentioned previously, the primary free parameters in our analysis are $\xi$, $\phi_{min}$ ($\varphi_{\min}$) and e-folding number $N_{inf}$. 
Once these values are fixed, we can find $g_X$ by the inflationary analysis, and then the $Z^\prime$ boson mass 
  ($M_{Z^\prime}=2 g_X \varphi_{min}$) and the reheating temperature from Eq.~(\ref{eq:k=aH}).

\begin{figure}[t]
\begin{center}
\includegraphics[height=0.4\textwidth]{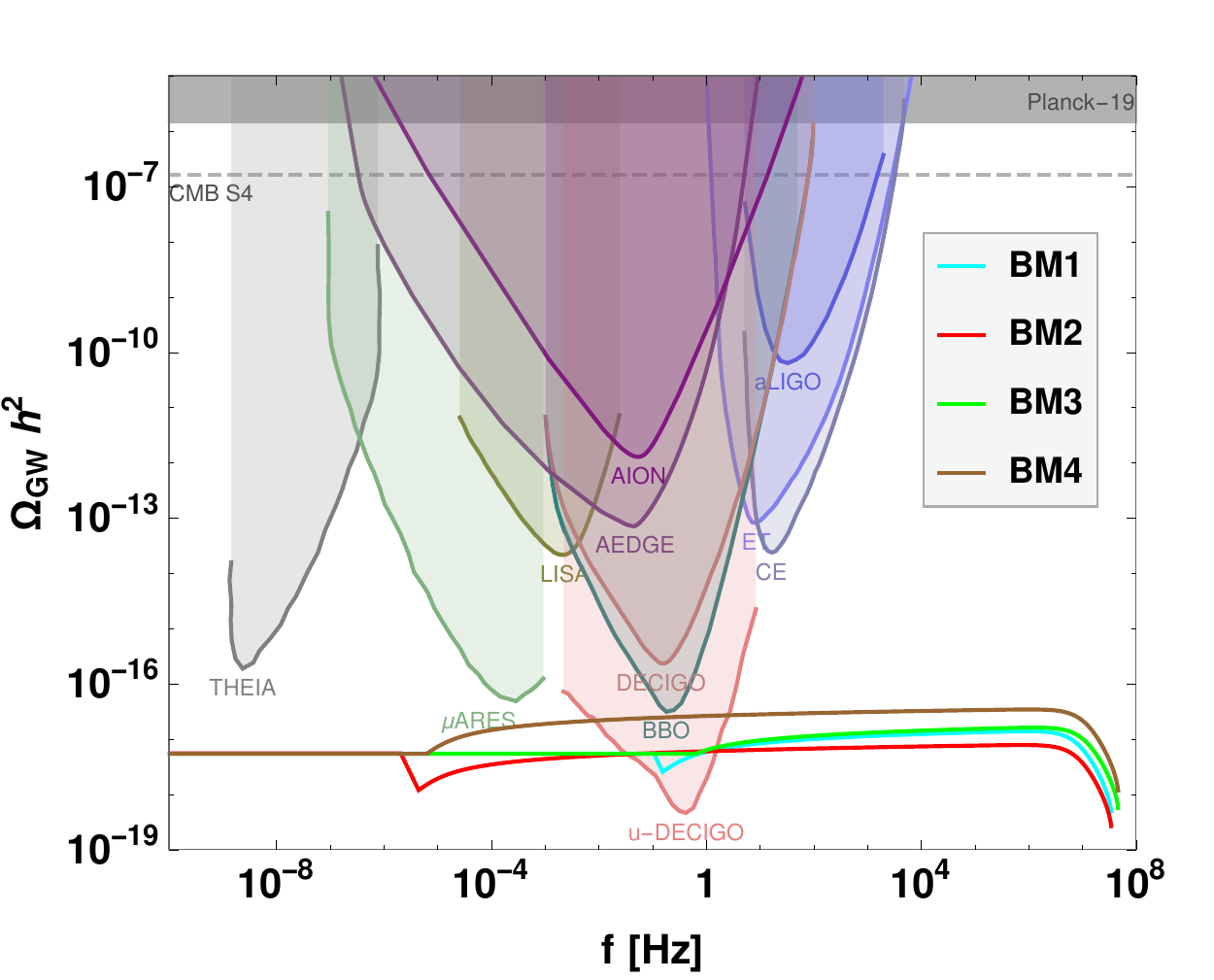}
\includegraphics[height=0.4\textwidth]{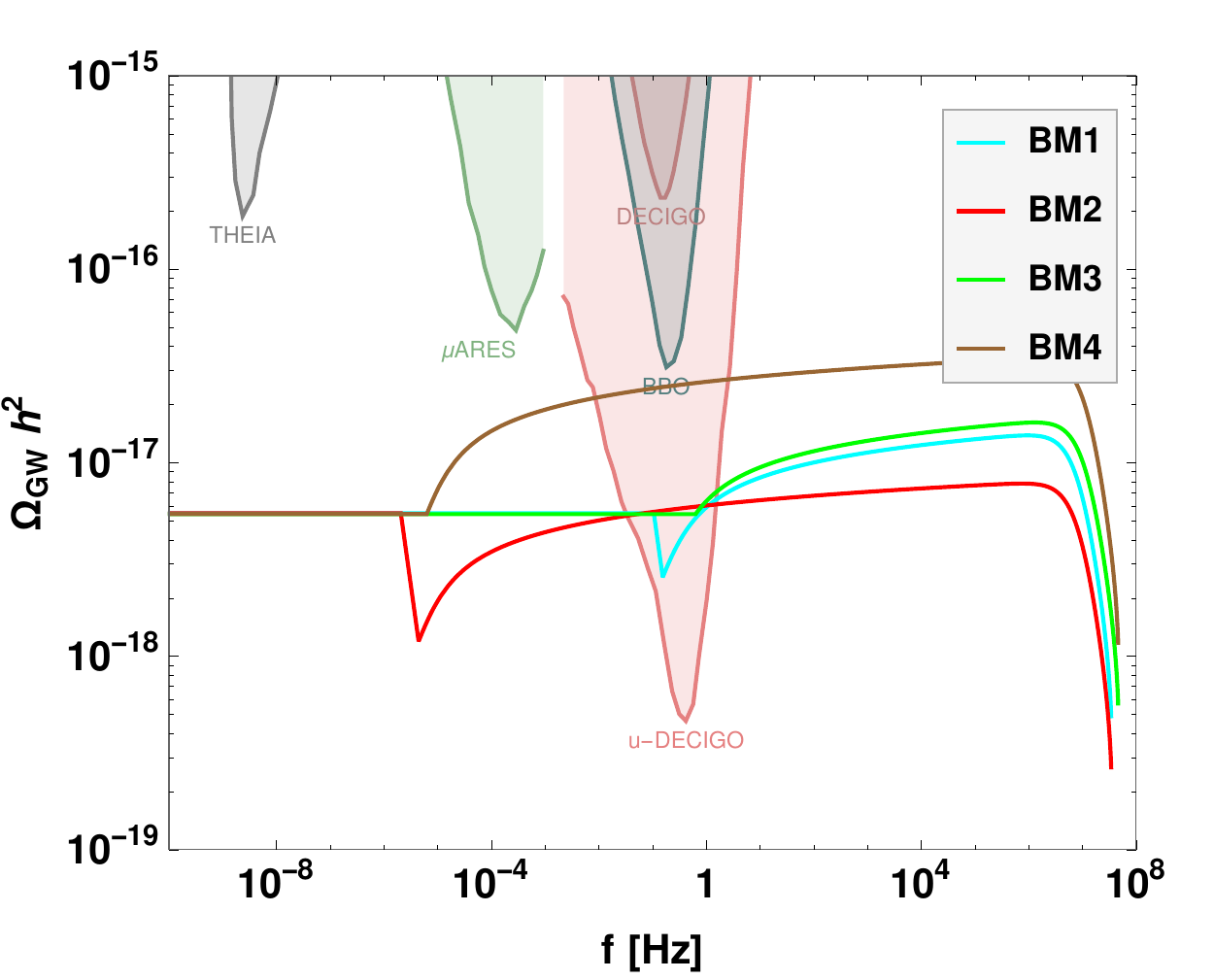}
	\caption {\it GW spectrum shown for the various benchmark points listed on TABLE \ref{t1},
	along with the prospective search reach of various planned/proposed GW detection experiments. We have shown the sensitivity curves from  LISA~\cite{LISA:2017pwj, Baker:2019nia}, BBO~\cite{Crowder:2005nr,Corbin:2005ny,Harry:2006fi}, DECIGO~\cite{Seto:2001qf,Kawamura:2006up,Yagi:2011wg}, CE~\cite{LIGOScientific:2016wof,Reitze:2019iox}, ET~\cite{Punturo:2010zz, Hild:2010id,Sathyaprakash:2012jk, Maggiore:2019uih}, $\mu-$ARES~\cite{Sesana:2019vho}, LVK~\cite{Harry:2010zz,LIGOScientific:2014pky,VIRGO:2014yos,LIGOScientific:2019lzm}.
 }
\label{fig:potinf2}
\end{center}
\end{figure}

In Fig.~\ref{fig:potinf2}, we show the resultant GW spectrum for the benchmark points listed in TABLE \ref{t1}.
The kination-like era induced by the Coleman-Weinberg potential in the minimal $U(1)_X$ model 
  modulates the GW spectrum from the flat spectrum in certain ranges of frequencies. 
As discussed before and also shown in Fig. \ref{fig:potinf}, $w_{avg}(t)$ grows slowly from $\sim 1/3$ to $\gtrsim .4$ during the phase $a_{end}$ to $a_{EMD}$ due to the radiatively corrected inflaton potential. The tensor modes (from $k_{end}=a_{end}H_{end}$ to the right to $k_{EMD}=a_{EMD}H_{EMD}$ at the left) reentering the Hubble sphere during this phase gets modulated imprinting a higher slope to the modes that reenter toward the end of the phase (as $w_{avg}(t)$ grows during this time). This describes the curved profile of GW spectral energy densities in Fig.~\ref{fig:potinf2}. After the phase the EMD phase with $w_{avg}=0$ continues till $a_{rh}$ giving rise to a negative slope in the GW spectrum for modes reentering during the phase. After the reheating is complete at $a_{rh}$, $w_{avg}$ becomes $0$, hence explaining the flat spectrum to the left of $k_{rh}=a_{rh}H_{rh}$. Note that near $k\lesssim k_{end}$ there is a sharp fall. This is due to the non-minimal coupling of $\phi$ to gravity which flattens $V_{\rm eff}(\varphi)$ at large $\phi$ ($\xi \phi^2 \gtrsim 1$), hence decreasing $w_{avg}$ from $1/3$ towards $-1$ as $\phi$ increases in the SAC formalism.
Benchmark 4 is particularly interesting in the viewpoint for the complementarity between the GW detections 
  and the high-energy collider search for the $Z^\prime$ boson: 
  the predicted GW spectrum is enhanced and well-overlaps with u-DECIGO, and 
  the $Z^\prime$ boson mass of about 1.6 TeV has been searched at the LHC experiment.

\subsection{$Z^\prime$ boson search at the LHC}

\begin{figure}[t]
\begin{center}
\includegraphics[height=0.45\textwidth]{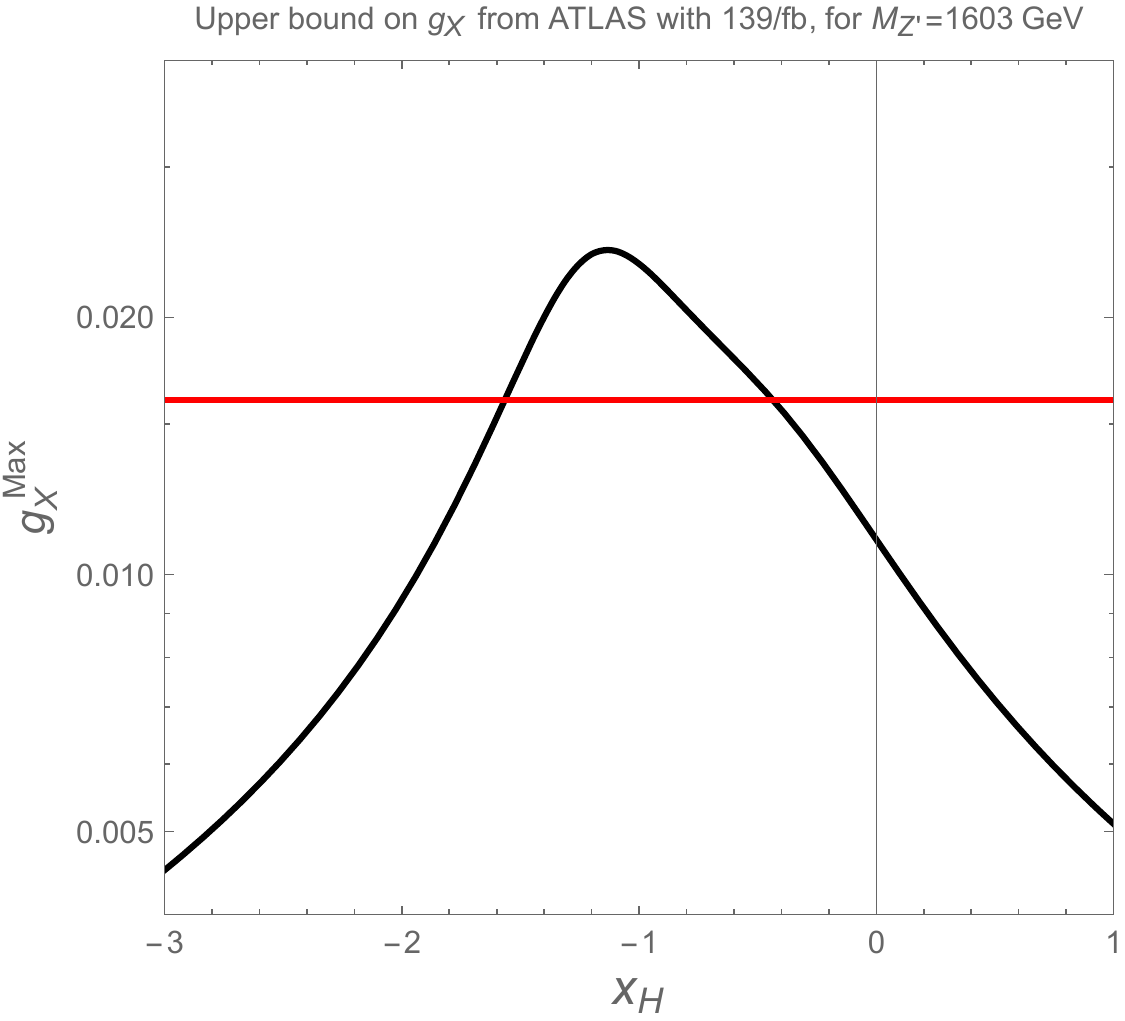}
	\caption{\it Upper bound on $g_X$ as a function of $x_H$ from the ATLAS 2019 result (black solid curve),
	along with $g_X=0.0160$ from BM 4 in TABLE \ref{t1} (red horizontal line).
		 }
\label{fig:potinf3}
\end{center}
\end{figure}

The ATLAS and the CMS Collaborations have been searching for $Z^\prime$ boson resonance at the LHC Run-2 with $\sqrt{s}=13$ TeV. 
The most stringent bounds on the $Z^\prime$ boson production cross section have been obtained by using the dilepton final state 
   $\ell^{+} \ell^{-}=e^+ e^-/\mu^+ \mu^-$.  
For the sequential SM $Z^\prime$  ($Z^\prime_{SSM}$) \cite{Barger:1980ti}, 
  the ATLAS and the CMS Collaborations have reported the lower mass bound on $Z^\prime$ as 
  $M_{Z^\prime_{SSM}} > 5.1$ TeV \cite{ATLAS:2019erb} and 
  $M_{Z^\prime_{SSM}} > 5.15$ TeV \cite{CMS:2019tbu}, respectively. 
In the following, we interpret the LHC Run-2 result into the U(1)$_X$ $Z^\prime$ boson case.

Since the ATLAS and the CMS results are consistent with each other, we interpret the ATLAS result into our case. 
We use the narrow width approximation to evaluate the $Z^\prime$ boson production cross section. 
The total $Z^\prime$ boson production cross section is given by
\begin{eqnarray}
	\sigma (p p \to Z^\prime)
		&=& 2 \sum_{q, \bar{q}} \int d x \int d y \: f_{q}(x, Q) \: f_{\bar{q}} (y, Q)  
			\: \hat{\sigma} (\hat{s}),
	\label{Eq:Zp_CrossSection}
\end{eqnarray}
  where $Q$ is the factorization scale (which we fixed $Q= M_{Z^\prime}$, for simplicity),  
   $\hat{s}= x y s$ is the invariant mass squared of the colliding partons (quarks) 
   with the center-of-mass energy $\sqrt{s}=13$ TeV of the LHC Run-2, 
   and $f_q$ ($f_{\bar{q}}$) is the parton distribution function (PDF) for quark (anti-quark).
In the narrow width approximation, the $Z^\prime$ boson production cross section at the parton level is expressed as
\begin{eqnarray}
	\hat{\sigma} (\hat{s})
		&=& \frac{4 \pi^2}{3} \frac{\Gamma (Z^\prime \to q \bar{q})}{M_{Z^\prime}} \: \delta (\hat{s} - M_{Z^\prime}^2) ,
	\label{Eq:Zp_CrossSection_atParton}
\end{eqnarray}
where $\Gamma(Z^\prime \to q \bar{q})$ is the $Z^\prime$ boson partial decay width to $q \bar{q}$. 
For the up-type and down-type quarks, respectively, 
\begin{eqnarray}
	\Gamma (Z^\prime \to u \bar{u})
		&=& \frac{g_X^2}{72 \pi} ( 2 + 10 x_H + 17 x_H^2) M_{Z^\prime}, \nonumber \\
	\Gamma (Z^\prime \to d \bar{d})
		&=& \frac{g_X^2}{72 \pi} ( 2 - 2 x_H + 5 x_H^2) M_{Z^\prime}.
	\label{Eq:Zp_DecayWidth}
\end{eqnarray}
We then simply express the dilepton production cross section as
\begin{eqnarray}
  \sigma( pp \to Z^\prime \to \ell^+ \ell^-) 
  &\simeq& \sigma( pp \to Z^\prime) \times {\rm BR}(Z^\prime \to \ell^+ \ell^-)
\end{eqnarray}
with the branching ratio to $Z^\prime \to \ell^+ \ell^-$, where $\ell=e$ or $\mu$. 
In our numerical analysis, we employ CTEQ6L~\cite{Pumplin:2002vw} for the PDF.

In the minimal $U(1)_X$ model,  the dilepton production cross section depends on three free parameters, 
  $g_X$, $M_{Z^\prime}$ and $x_H$. 
Setting $M_{Z^\prime}=1603$ GeV from BM 4 in TABLE \ref{t1}, 
  we interpret the ATLAS bound into an upper bound on $g_X$ as a function of $x_H$. 
Fig.~{fig:potinf3} shows our result, along with $g_X=0.0160$ from BM 4 in TABLE \ref{t1}. 
The ATLAS result constraints $x_H$ to be within the range of $x_H$ to be $-1.57 < x_H < -0.434$.

\medskip

\section{Conclusion and Discussion}
\label{conclusion}
It is known that if a kination-like era with a stiff equation of state, $w>1/3$, is realized after inflation, 
  the irreducible stochastic gravitational wave background from inflation can be substantially enhanced/modulated 
  from the conventional flat spectrum. 
Such a distinct GW spectrum could be detectable by next-generation GW interferometer missions, such as U-DECIGO. As explained in the last paragraph of \ref{mod}, this radiatively corrected inflaton potential gives rise to a uniquely curved shape in the GW spectrum, which is starkly different from other stiff e.o.s scenarios where $w_{avg}$ is mostly constant. This gives us a possibility to disentangle such scenarios with radiatively corrected potentials in future powerful GW observations. 

Based on the minimal $U(1)_X$-extended Standard Model, which is a well-motivated Beyond the Standard Model 
  from various points of view, we have explored cosmic inflation 
  by identifying the $U(1)_X$ Higgs field with the inflaton. 
Assuming the classically conformal $U(1)_X$ Higgs sector, we have analyzed the inflaton dynamics 
  during its oscillatory phase after inflation along the Coleman-Weinberg effective potential. 
We have found that the Coleman-Weinberg potential which is steeper than $\phi^4$ potential 
  for a large inflaton amplitude drives the kination-like era in the early universe 
  until the inflaton gets trapped in a potential minimum and the early matter-dominated era begins.

The inflaton potential is controlled by only two free parameters, non-minimal inflaton coupling with gravity $\xi$ 
  and the inflaton/Higgs VEV, $\phi_{min}$, developed by the radiative $U(1)_X$ symmetry breaking, 
  under the constraints on the inflationary predictions. 
Once the two parameters are fixed, the $U(1)_X$ gauge coupling ($g_X$) and the $Z^\prime$ boson mass ($M_{Z^\prime}$)
  are fixed from the inflationary analysis with a fixed e-folding number. 
The reheating temperature is also fixed in a consistent manner. 
We have provided four benchmark points in TABLE \ref{t1}. 
The predicted gravitational wave spectrum corresponding to each benchmark point is depicted in Fig.~\ref{fig:potinf2}.
We can see that the kination-like era substantially modulates/enhances the gravitational wave spectrum. 
The benchmark points with $M_{Z^\prime}={\cal O}$(1 TeV) are particularly interesting 
   since such a TeV scale $Z^\prime$ boson has been searched for at the LHC. 
This fact leads us an idea of complementarity between the gravitational wave detections and 
   $Z^\prime$ boson resonance searches at high-energy collider experiments. 
For the BM 4 in TABLE \ref{t1}, we have demonstrated this this complementarity (see Figs.~\ref{fig:potinf2} and \ref{fig:potinf3}).

 With the initiative known as the ``UHF-GW Initiative \cite{Aggarwal:2020olq}, there is ever growing interests in designing and planning to detect very high frequency GW. This has lead to several new ideas for detection techniques, e.g., \cite{Ballantini:2005am,Arvanitaki:2012cn,Ejlli:2019bqj,Aggarwal:2020umq,LSD:2022mpz,Berlin:2021txa,Berlin:2022hfx,Berlin:2023grv,Goryachev:2021zzn,Goryachev:2014yra,Campbell:2023qbf,Sorge:2023nax,Tobar:2023ksi,Carney:2023nzz,Domcke:2022rgu,Domcke:2023bat,Bringmann:2023gba,Vacalis:2023gdz,Liu:2023mll,Ito:2019wcb,Ito:2020wxi,Ito:2022rxn,Ito:2023bnu}, 
  although going beyond the BBN dark radiation bound still remains very challenging from detection sensitivity point of view. 
As we can see from Fig.~\ref{fig:potinf2}, the predicted gravitational wave spectrum is significantly enhanced in the high frequency region 
  for $M_{Z^\prime}$ exceeding the energy scale of high energy colliders (see BM 1 and BM 3). 
Our $U(1)_X$ scenario with realizing the kination-like era provides a concrete science case for UHF-GW detectors, 
  which could probe particle physics at energy scales many orders of magnitude beyond the reach of future high energy colliders.

\section{Acknowledgements}
The work of N.O. is supported in part by the United States Department of Energy Grant Nos.~DE-SC0012447 and DESC0023713. A.P. thanks the Indo-French Centre for the Promotion of Advanced Research for supporting the postdoctoral fellowship through the proposal 6704-4 under the Collaborative Scientific Research Programme.



\bibliographystyle{utphysII}

\bibliography{ref2}
\end{document}